\documentclass[aps,pra,twocolumn,superscriptaddress]{revtex4}

\usepackage{psfrag}
\usepackage{subfigure}
\usepackage[dvips]{graphicx}
\usepackage{amsmath}

\begin{document}


\title{Fast shuttling of ions in a scalable Penning trap array}

\author{D.~R.~Crick}
\author{S.~Donnellan}
\affiliation{Department of Physics, Imperial College, London SW7 2AZ, U.K.}
\author{S.~Ananthamurthy}
\affiliation{Department of Physics, Bangalore University, Bangalore 560056, India}
\author{R.~C.~Thompson}
\author{D.~M.~Segal}
\affiliation{Department of Physics, Imperial College, London SW7 2AZ, U.K.}

\date{\today}

\begin{abstract}
We report on the design and testing of an array of Penning ion traps made from printed circuit board.  The system enables fast shuttling of ions from one trapping zone to another, which could be of use in quantum information processing.  We describe simulations carried out to determine the optimal potentials to be applied to the trap electrodes for enabling this movement.  The results of a preliminary experiment with a cloud of laser cooled calcium ions demonstrate a round-trip shuttling efficiency of up to 75\%.
\end{abstract}


\maketitle

\section{Introduction}

Laser cooled trapped ions constitute one of the most advanced experimental approaches to quantum information processing (QIP) (see ref~\cite{haeffner} and references therein). Many of the key building blocks of QIP have been demonstrated using chains of ions in linear radiofrequency traps; two-ion quantum gates have been performed~\cite{blattcz,leibfried1} and entanglement of up to 8 qubits has been demonstrated~\cite{wineland:6ion, blatt:8ion}. However, it is widely assumed that scaling the ion trap system to larger numbers of qubits will require arrays of micro-traps in which ion-qubits are shuttled between trapping zones~\cite{kielpinski}. A number of the key elements of this approach have been demonstrated; ions have been shuttled between different trapping zones in multiple trap architectures~\cite{wineland_teleport} whilst retaining internal state coherence and ion-pairs have been split with low loss of coherence~\cite{wineland_shuttle}. Since a universal quantum computer relies on the ability to perform quantum gates between any pair of qubits in the quantum register the ability to sort ions is imperative: Assuming nearest neighbour interactions only~\footnote{Schemes exist for which nearest neighbour interactions are not required. See~\cite{toschek}.} it must be possible to remove a given ion from a chain and bring it into contact with a different ion. This fundamentally requires ions to be moved in two dimensions, or more specifically that an ion can be moved around a corner. This procedure has already been demonstrated in a T-junction radiofrequency trap array albeit with significant heating~\cite{move_round_corner}.

As an alternative to ions held in radiofrequency traps, ions contained in Penning traps, which use only static electric and magnetic fields to provide confinement, have been proposed for applications in quantum simulation~\cite{porras_cirac} and quantum information processing~\cite{padtrapjmo, kingston}.  Single ions and pairs of ions have been imaged and aligned along the the axis of a Penning trap~\cite{small_crystals}.  Arrays of Penning micro-traps could also be used, although the operation of very small Penning traps has yet to be demonstrated.  Moving ions in two dimensions in such trap arrays will also be a requirement.  It should be possible to move ions along the magnetic field axis of a Penning trap using the same kind of techniques for ion shuttling that have been pioneered in radiofrequency traps, since the magnetic field has no effect on motion in this direction. Motion perpendicular to the magnetic field axis is however inherently different in a Penning trap due to the ${\mathbf v} \times {\mathbf B}$ component of the Lorentz force. A method for overcoming this complication has been presented elsewhere~\cite{padtrapjmo}. It involves exploiting the cycloid motion of a charged particle in crossed electric and magnetic fields to shuttle an ion from one trapping zone to another by the application of a pulsed electric potential. The movement can be rather fast since it occurs in approximately one cyclotron period ($\sim$2.5~$\mu$s for Ca$^+$ in a $B = 1$~tesla magnetic field). Furthermore, an ion that starts at rest in one trapping zone automatically comes back to rest in the target zone. This paper reports the first realisation of a multiple Penning trap architecture capable of performing this transfer of ions  between trapping zones separated in a plane perpendicular to the magnetic field axis. The controlled movement of clouds of ions between trapping zones is presented.

\section{Design and Manufacturing}

\subsection{Boards}

The Penning trap array is made from printed circuit board.  FR4 (Flame Retardant 4) board is a composite of an epoxy resin reinforced with a woven fibreglass mat.  The vacuum properties of FR4 board have been studied, and the outgassing rate found to be low enough for UHV applications~\cite{FR4_vacuum}.  Two boards are mounted facing each other, with the magnetic field direction normal to the board surface.  Copper pads on the board surfaces act as the trap electrodes.  Close to the centre of the trap, the potential is very similar to an ideal Penning trap.

\begin{figure}[htbp]
 \centering
\psfrag{A}{$A$}
\psfrag{B}{$B$}
\psfrag{C}{$C$}
\psfrag{D}{$D$}
\psfrag{E}{$E$}
\psfrag{33.5}{$33.5$}
\psfrag{41}{$41$}
\includegraphics[width=0.35\textwidth]{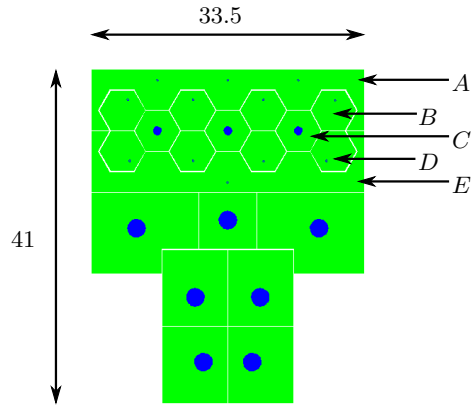}
\caption{Board design.  Dimensions in mm.}
\label{fig:padtrap_electrodes}
\end{figure}

The board design is shown in figure~\ref{fig:padtrap_electrodes}.  Almost all of the board area is covered with copper so that any effects due to charge buildup on the insulating surface are kept to a minimum.  The electrodes are arranged in rows of hexagons.  The hexagonal pads in each row are electrically connected together with copper vias running through the board on to tracks on the rear side.  Three hexagonal pads on each board, labelled $C$ in figure~\ref{fig:padtrap_electrodes}, act as the endcap electrodes for three trapping zones.  Each of these has a 1~mm diameter hole at the centre.  Above and below the endcap electrodes are rows of pads labelled $B$ and $D$.  The remaining electrodes are labelled $A$ and $E$.  The trap is made from two parallel boards facing each other with a gap of 5.0~mm.  The design on each is a mirror image of the opposing board.  In normal operation the $A$, $B$, $D$, $E$ electrodes on the pair of boards act as the ring electrodes for the three traps.  There are seven 2~mm diameter holes below the trapping regions.  Five  of these are used to make connections to the five sets of electrodes.  These are connected to the vacuum feedthrough via wires bolted onto the rear of the boards.  The outermost two holes are required to mechanically attach the trap to the rest of the superstructure.

A  CNC  Milling/Drilling machine was used to drill the holes in the board, which were then  electroplated with copper using an LPKF through-hole plating system.  This process connects the front and rear sides of the board through the holes.  The CNC machine was then used to mill away gaps between the electrodes.  The gaps are approximately 200~$\mu$m wide.  

\subsection{Superstructure and Vacuum Chamber}\label{sec:superstructure}

\begin{figure}[htbp]
\flushleft
\hspace{0.2cm}
\psfrag{Lens holder}{\footnotesize{Lens holder}}
\psfrag{Calcium oven}{\footnotesize{Calcium oven}}
\psfrag{Stainless steel support}{\footnotesize{Stainless steel support}}
\psfrag{Copper support}{\footnotesize{Copper support}}
\psfrag{Feedthrough flange}{\footnotesize{Feedthrough flange}}
\psfrag{PCB}{\footnotesize{PCB}}
\psfrag{Hot cathode}{\footnotesize{Hot cathode}}
\psfrag{filament}{\footnotesize{filament}}
\includegraphics[width=0.4\textwidth]{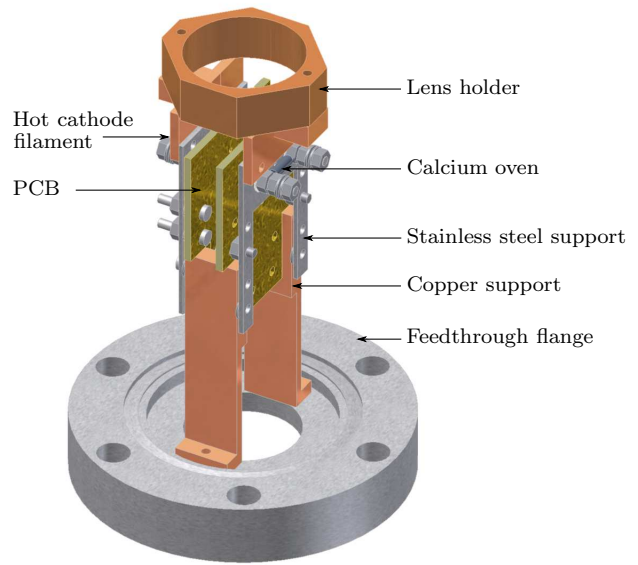}
\caption{Diagram of the trap.  Some parts have been removed for clarity.}
\label{fig:pcb_trap_photos}
\end{figure} 

A diagram of the assembled trap is shown in figure~\ref{fig:pcb_trap_photos}.  A pair of copper supports are bolted to a CF40 flange with an 11 pin electrical feedthrough.  Stainless steel plates are attached to the copper legs.  The steel plates hold the trap boards, a filament and an oven (used for loading the trap), and the lens holder.  All of the screws, metal washers and nuts are size M2 non-magnetic stainless steel.

The boards are held 5.0~mm apart by stainless steel spacers.  The spacers also provide an electrical connection between the pair of boards.  In the final iteration of the trap, the spacer which would connect electrode $C$ on each board was not used so that the endcaps on each side could be connected independently.  The spacer connecting electrode $E$ is also absent so as to minimise   the laser scatter.  Kapton coated wires are used to connect the electrodes to the feedthrough pins underneath the trap.  Kapton wires are also used to connect to the oven and the filament.

\begin{figure}[htbp]
\centering
\includegraphics[width=0.4\textwidth]{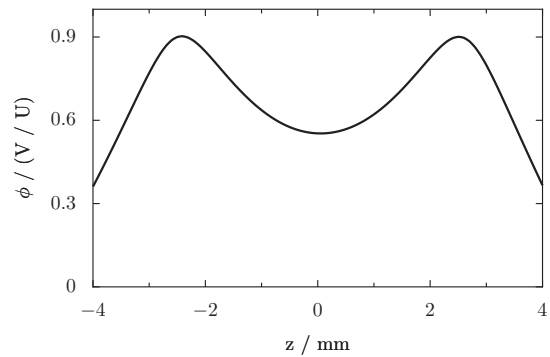}
\caption{Potential along the axis of the trap.  Simulation grid size was 0.1~mm.}
\label{fig:pot_alongz}
\end{figure}

The 5.0~mm spacing between the pair of boards was chosen to make the trapping potential as harmonic as possible close to the centre of the trap (within 1~mm).  The simulated potential along the central axis is shown in figure~\ref{fig:pot_alongz}.

Calcium ions are created by electron impact ionisation which is facilitated by counterpropagating electron and atom beams.   A 1~mm diameter hole in the steel plate, midway between the screws, is concentric with a 1~mm diameter hole in each of the trap boards.  These holes allow the atomic beam and electron beam to pass through the centre of the trap. The positioning of the filament is chosen to allow electrons to propagate along the magnetic field lines.

The lens  is a 40~mm focal length singlet lens and is positioned 20~mm above the centre of the middle trapping zone for maximum light gathering efficiency.    A piece of constantan foil with a 6~mm diameter hole is used as a baffle above the trap to block some of the scattered light coming from areas not directly underneath the lens.

\section{Simulations}\label{sec:hopping_simulations}

To shuttle ions from one trap to another, the voltages on the various electrodes are briefly switched from their normal value for trapping to new values for shuttling.  By applying ascending (or descending) voltages to $A$, $B$, $C$, $D$ and $E$, an approximately linear electric field can be produced, perpendicular to the magnetic field direction.  When the linear electric field is applied, an ion initially at rest will start to move along this direction but be pulled round in a cycloid loop by the $B$ field.  By appropriately choosing the magnitude of the electric field, the size of the cycloid loop will be the same as the spacing between adjacent traps.  By applying these voltages for the correct amount of time, an ion will move out of one trap and come to rest at the centre of the next trap.  Applying the opposite voltages for a similar duration will cause an ion to move between traps in the opposite direction.

\begin{figure}[htbp]
\centering
\psfrag{V / volt}{\LARGE{$\phi$ / V}}
\psfrag{x / mm}{\LARGE{$x$ / mm}}
\psfrag{y / mm}{\LARGE{$y$ / mm}}
\psfrag{z / mm}{\LARGE{$z$ / mm}}
\hspace{0cm} \mbox{\subfigure[Trapping potential]{ \includegraphics[angle=270,width=0.4\textwidth]{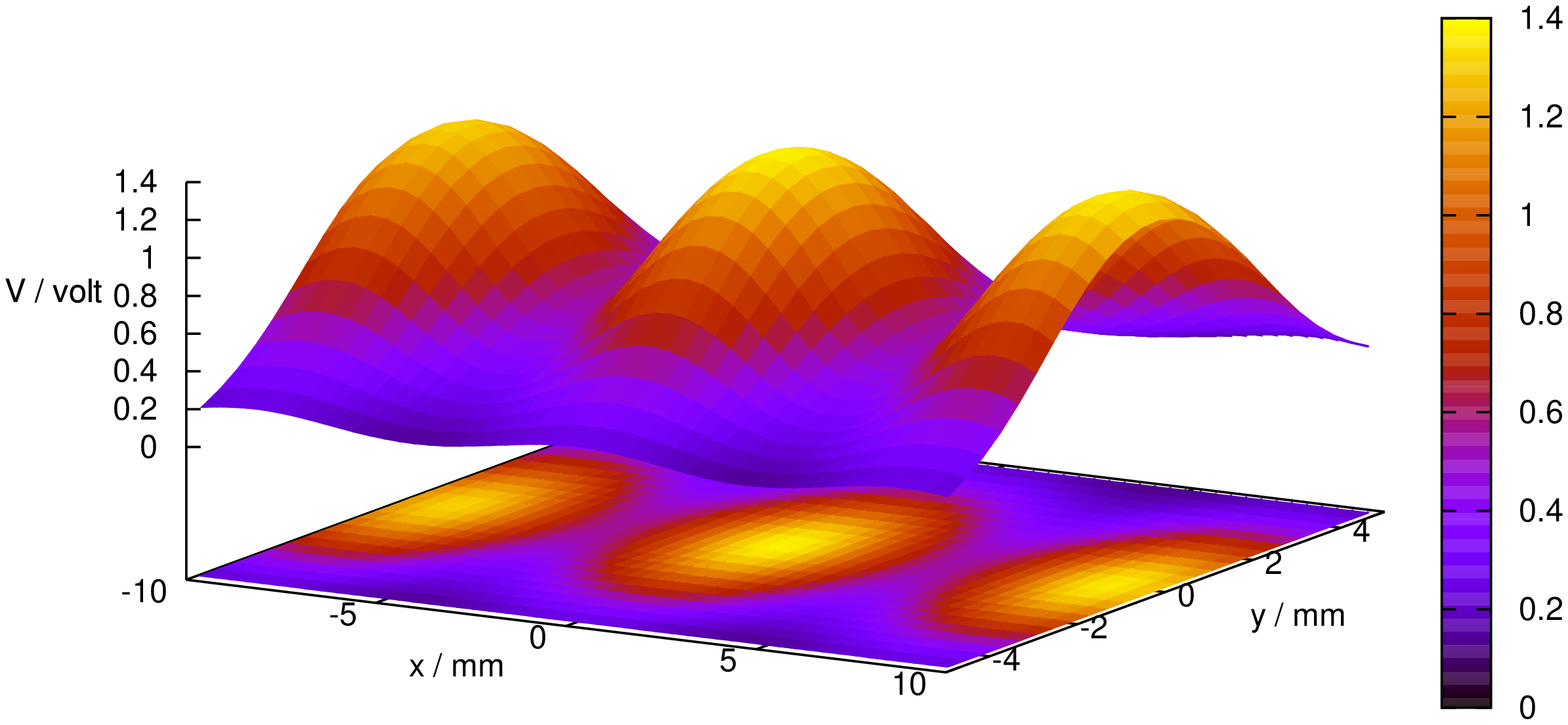}}}


\mbox{\subfigure[Shuttling potential]{ \includegraphics[angle=270,width=0.4\textwidth]{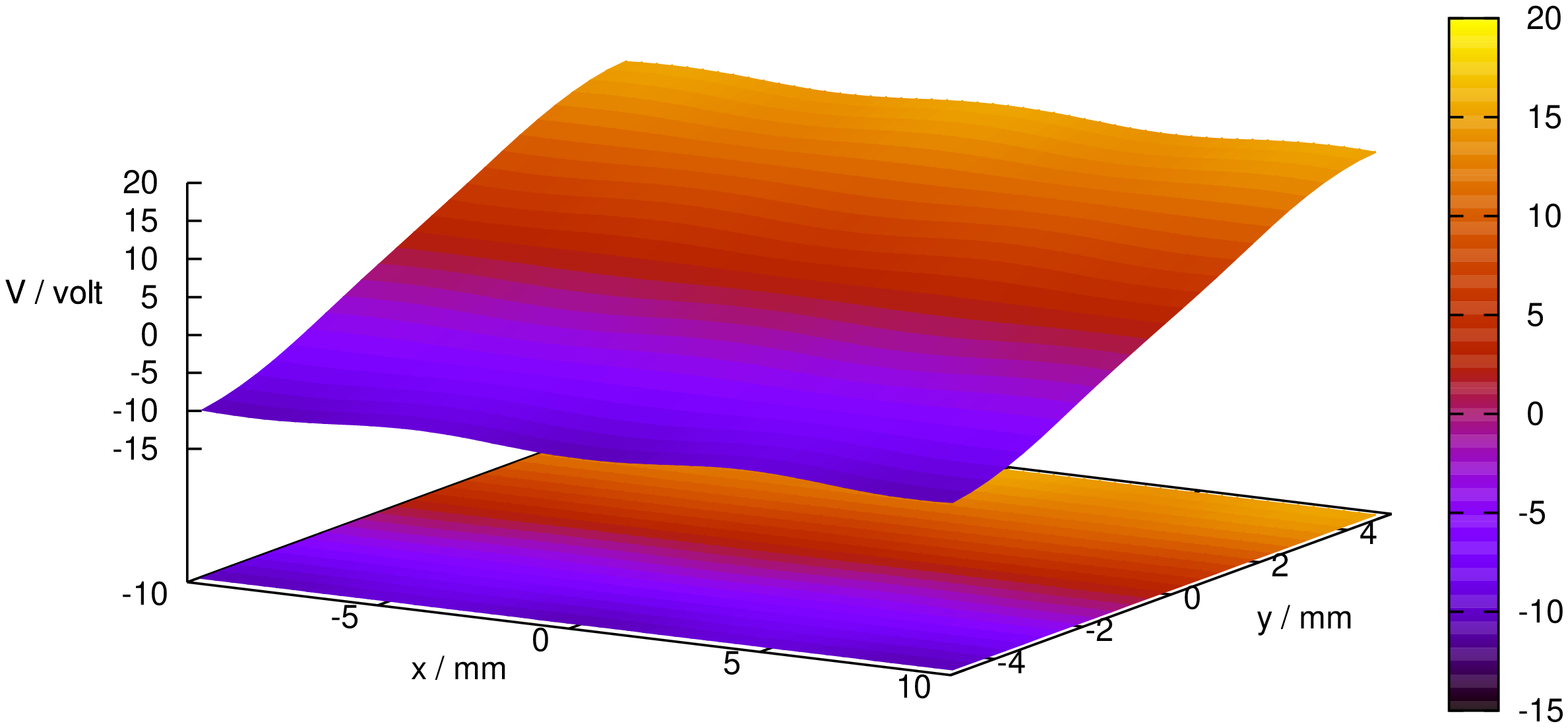} }}

\caption[Electrostatic potential in $xy$ plane]{Simulated electrostatic potentials in the $xy$ plane directly between the two trap boards.  Note the change in scale of the vertical axis.}
\label{fig:pcb_potential_maps}
\end{figure} 

\begin{figure}[htbp]
\centering
\includegraphics[width=0.4\textwidth]{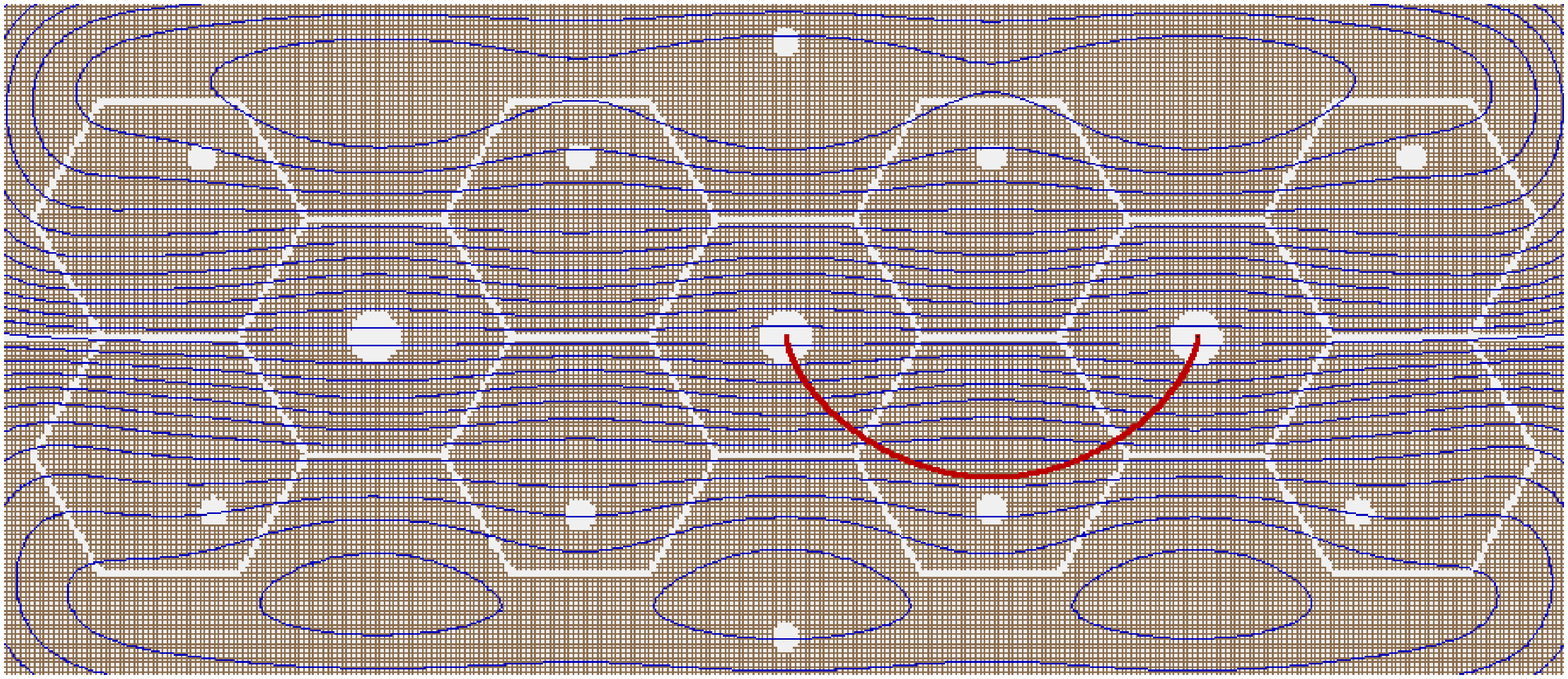}
\caption[Ion shuttling trajectory]{Example of an ion trajectory while moving from one trap to the next.  Lines of equipotential in the $xy$ plane are shown in blue.}
\label{fig:jumptraj}
\end{figure}

Figure~\ref{fig:pcb_potential_maps} shows the electrostatic potential in the plane directly between the pair of trap boards, when the voltages are set to `trap mode' and to `shuttle mode' respectively.  A typical trapping potential is formed when the voltages on the electrodes are $(V_A, V_B, V_C, V_D, V_E) = (0,0,2.5,0,0)$ volt.  Note from figure~\ref{fig:pcb_potential_maps}(a), that even though the electrodes are hexagonal, the shape of the potential close to the centre of the trap is almost perfectly circular.  The potential in figure~\ref{fig:pcb_potential_maps}(b) is produced when $(V_A, V_B, V_C, V_D, V_E) = (20,12.5,2.5,-7.5,-15)$ volt.  Figure~\ref{fig:jumptraj} shows an example of the trajectory of an ion as it moves between a pair of trapping regions.

In the simulations the electrode voltages, time durations, initial conditions, etc., can be made unrealistically perfect.  To learn roughly how sensitive the ion shuttling system is to experimental imperfections, various parameters were changed and the effect on a simulated ion was observed.  

\begin{figure}[htbp]
\centering
\includegraphics[width=0.4\textwidth]{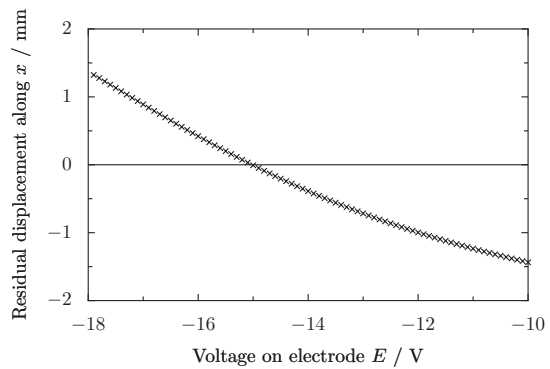}
\caption{Residual position of an ion relative to the centre of the second trap, immediately after shuttling.  $B = 0.9$~tesla.}
\label{fig:displacement_vs_voltage}
\end{figure}

The points in figure~\ref{fig:displacement_vs_voltage} show how far an ion misses the centre of the target trap if one of the voltages is not correct.  The ion begins at rest at the centre of the first trap.  Electrodes $A$, $B$, $C$ and $D$ were set to 20, 12.5, 2.5 and -7.5~V respectively.  Since the ion passes closest to electrode $E$ during most of its trajectory, a small change in $V_E$ has a bigger impact on the path of the ion than a small change in any of the other voltages does.  It can be seen that 1 volt of error in $V_E$ causes an ion to miss its target by $\sim$0.5~mm.

\begin{figure}[htbp]
\centering
\includegraphics[width=0.4\textwidth]{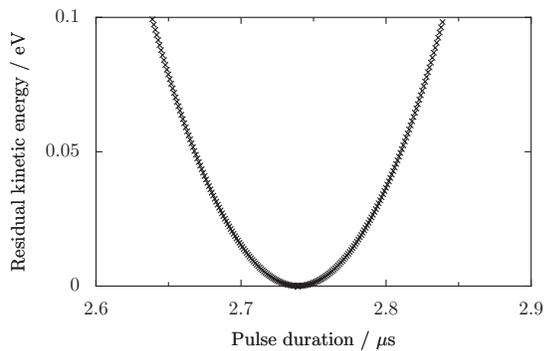}
\caption{Residual kinetic energy of an ion, immediately after shuttling an ion in a field of 0.9~tesla.}
\label{fig:ke_vs_timing}
\end{figure}

Figure~\ref{fig:ke_vs_timing} shows how an ion gains kinetic energy if the shuttling voltages are applied for too long or too short a time.  The voltages were set appropriately to move the ion to the centre of the second trap in a single cycloid loop.  It can be seen that the kinetic energy acquired by an ion varies quadratically with $t - t_{\textrm{ideal}}$, and reaches roughly 100~meV when the duration is wrong by 100~ns.

There is a range of possible sets of voltages which will cause an ion to move between the centres of adjacent traps.   When the ion is initially at rest and perfectly centred in the first trapping zone, all of these different voltage sets will produce good results.  However, if the ion begins at a position slightly offset from the trap centre  then this initial imperfection is amplified by different amounts depending on the specific set of voltages used. Since the ion moves perpendicular to the magnetic field, a displacement of the ion relative to the trap centre along $z$ is similar to a misalignment of the trap with the magnetic field. Finding the best possible parameters is a non-trivial problem.  It would involve finding a line in four dimensional voltage-space ($V_A, V_B, V_D, V_E$) for which an ion with ideal initial conditions reaches the centre of the target trap, and then finding the point on this line which produces a optimal final condition given a range of realistic initial conditions.  To simplify matters, we consider a set of voltages which are symmetric about electrode $C$: 
\begin{align}
\left(\begin{array}{c}
 V_A \\
 V_B \\
 V_C \\ 
 V_D \\ 
 V_E \end{array} \right)
 &= \left(\begin{array}{c}
 2.5+a \\
 2.5+\epsilon a \\
 2.5 \\
 2.5 - \epsilon a \\
 2.5 - a \end{array} \right) \label{eqn:symvoltages}
\end{align}
and also a set which is not symmetric:  
\begin{align}
\left(\begin{array}{c}
 V_A \\
 V_B \\
 V_C \\ 
 V_D \\ 
 V_E \end{array} \right)
 &= \left(\begin{array}{c}
 0.0 \\
 0.0 \\
 2.5 \\
 2.5 - \epsilon a \\
 2.5 - a \end{array} \right) \label{eqn:asymvoltages}
\end{align}
The asymmetric set has the advantage that only two electrodes need to be switched instead of four.  

\begin{figure}[htbp]
  \centering
    \mbox{
      \subfigure[Optimal voltage (symmetric)]{{\includegraphics[width=0.25\textwidth]{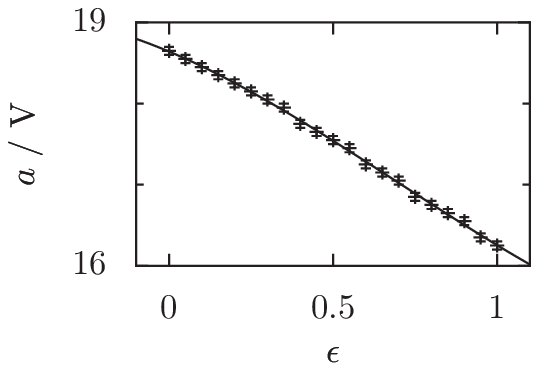}}} 
      \subfigure[Optimal voltage (asymmetric)]{{\includegraphics[width=0.25\textwidth]{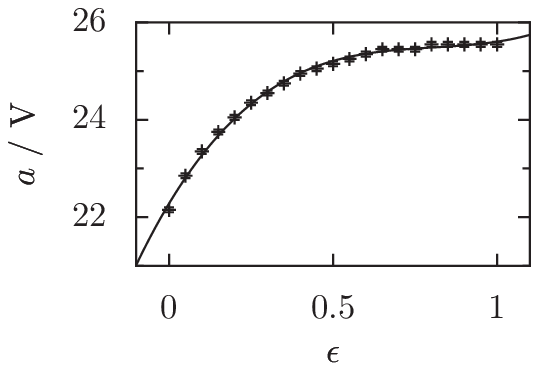}}}
      }
    \mbox{
      \subfigure[Optimal duration (symmetric)]{{\includegraphics[width=0.25\textwidth]{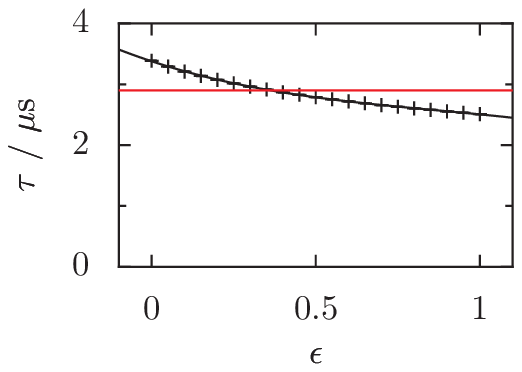}}} 
      \subfigure[Optimal duration (asymmetric)]{{\includegraphics[width=0.25\textwidth]{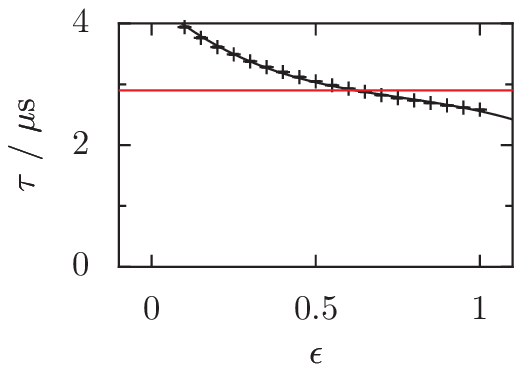}}}
     }
     \caption{Simulated optimal voltage and timing parameters for $B = 0.9$~tesla.  The true cyclotron period is also shown in (c) and (d).}
    \label{fig:best_a_values}
\end{figure}

The simulation is scanned over a range of values of $\epsilon$.  For each $\epsilon$, an optimal value for $a$ is found (along with an optimal pulse duration $\tau$), and then some measure of the final condition of the ion can be found for a particular initial condition.  The best value of $a$ is found by minimising the distance along $x$ between the ion and the centre of the target trap, immediately after shuttling. The best duration, $\tau$, is defined as the time between the application of the shuttling voltages and the next minimum in the ion's kinetic energy.  The best values of $a$ and of the pulse duration are shown in figure~\ref{fig:best_a_values}. 

\begin{figure}[htbp]
  \centering
    \mbox{
     \subfigure[RMS displacement (sym)]{{\includegraphics[width=0.2\textwidth]{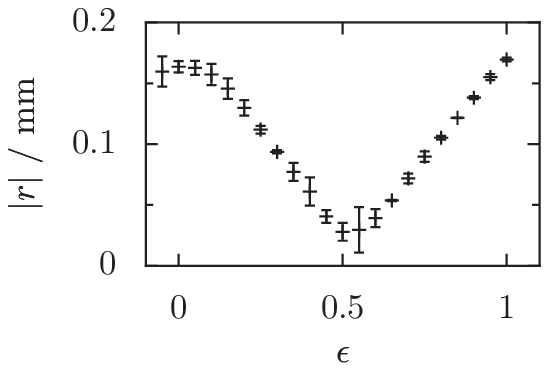}}} 
     \subfigure[RMS displacement (asym)]{{\includegraphics[width=0.2\textwidth]{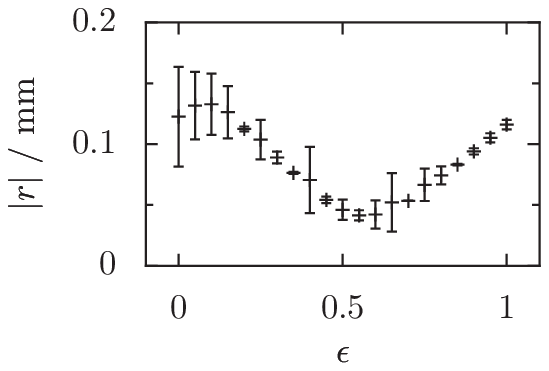}}}
      }
    \mbox{
      \subfigure[Velocity along $z$ (sym)]{{\includegraphics[width=0.2\textwidth]{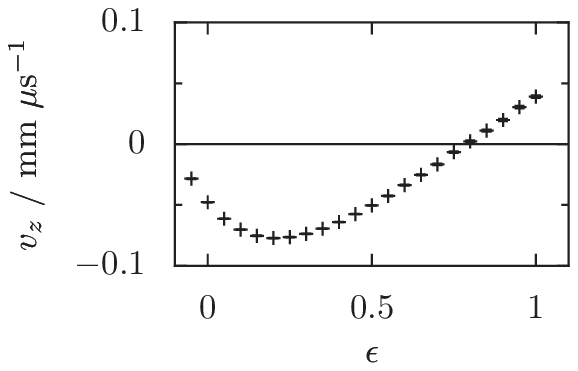}}} 
      \subfigure[Velocity along $z$ (asym)]{{\includegraphics[width=0.2\textwidth]{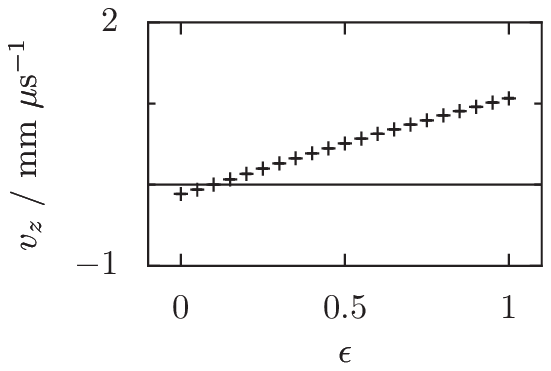}}}
     }
    \mbox{
      \subfigure[KE (sym)]{{\includegraphics[width=0.2\textwidth]{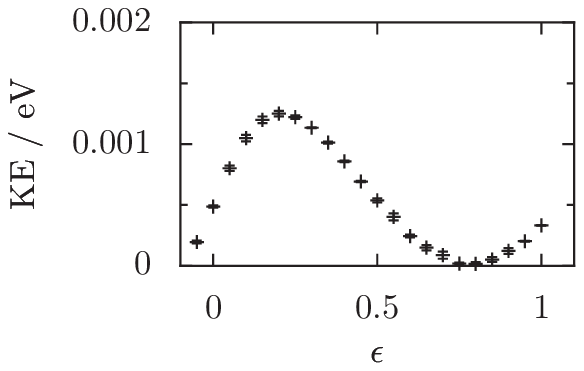}}}
      \subfigure[KE (asym)]{{\includegraphics[width=0.2\textwidth]{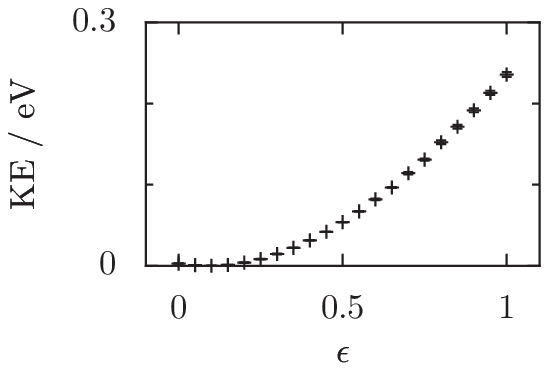}}} 
	}
    \mbox{
      \subfigure[Total energy (sym)]{{\includegraphics[width=0.2\textwidth]{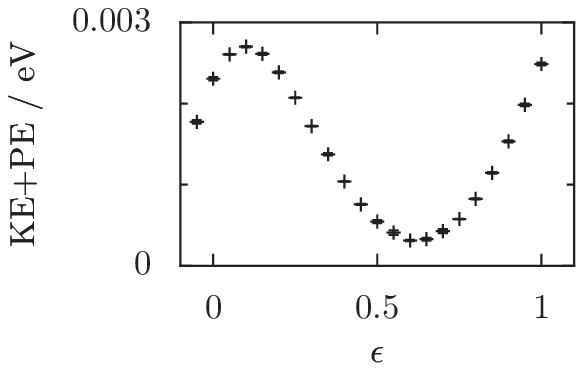}}} 
      \subfigure[Total energy (asym)]{{\includegraphics[width=0.2\textwidth]{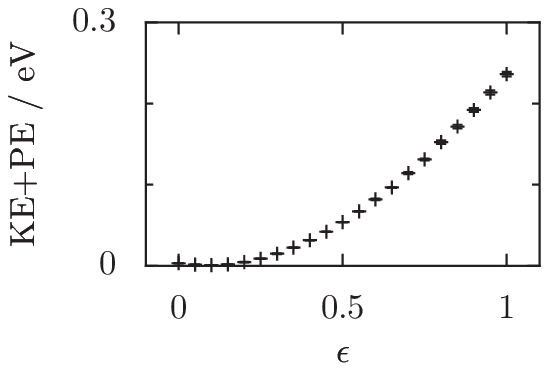}}}
     }
     \caption{Final conditions of an ion after shuttling for various voltage sets.  The ion was initially at rest and displaced 0.1~mm along $z$.  $B = 0.9$~tesla.}
    \label{fig:epsilon_scan}
\end{figure}

Figure~\ref{fig:epsilon_scan} show how different values of $\epsilon$ lead to different final conditions of an ion.  The initial condition is such that the ion is at rest but displaced by 0.1~mm along the axis of the first trap.  The final position is plotted, along with the final velocity, the final kinetic energy, and final total energy.  All of these measures should ideally be as small as possible.

It was found that the symmetric voltage set,~(\ref{eqn:symvoltages}), produced better results (lower final KE etc.) than the asymmetric set,~(\ref{eqn:asymvoltages}).  The asymmetric voltage sets cause the ion to gain significantly more kinetic energy.  It was also found that the best results were obtained when $\epsilon \approx 0.5$ -- i.e. a linear step down of the five electrode voltages.

An initial displacement of the ion along $x$ or $y$ has less effect on the outcome than an initial displacement along $z$.  Other simulations show that that for typical voltage sets there is an overall restoring force pushing ions towards $z = 0$.  Unfortunately, even with this restoring force present, any initial displacement or initial velocity will cause an ion to gain energy when moving between traps. An initial displacement of 0.1~mm along $z$ is roughly equivalent to a misalignment between the trap and the magnetic field of just half a degree.  Clearly this misalignment must be reduced as much as possible when attempting to shuttle ions in the real trap.

\section{Electronics}

In order to produce the voltage pulses needed to shuttle ions between traps, a system of high speed switching electronics was developed.  Each electrode (except for electrode $C$) needs to be switchable between 0~V (or close to 0~V) for trapping, a positive voltage of up to around 20~V for shuttling ions in one direction, and a negative voltage of similar magnitude for shuttling in the other direction.  The voltages need to be set precisely, and to remain stable over time.  The switching needs to occur on a timescale which is short compared to the length of the pulse.  The pulse length must be set precisely and must also remain stable over time.  There should also be a negligible relative time delay between the pulses on the various electrodes.

\begin{figure}[htbp]
\centering
\includegraphics[width=0.45\textwidth]{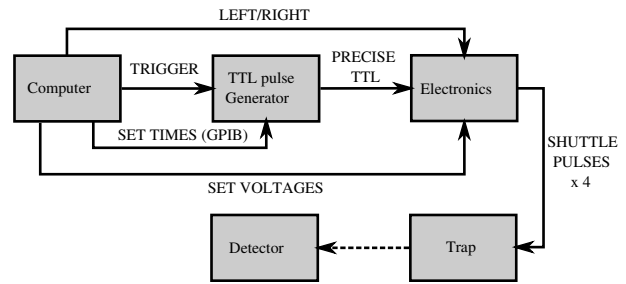}
\caption{Generalised layout of the system used for shuttling ions.}
\label{fig:hopping_electronics_general}
\end{figure}

To generate pulses of a precise duration, a commercial digital pulse generator (Stanford Reseach Systems DG535) is used to produce TTL pulses.  The TTL pulses are then converted into pulses of the required voltages.  This amplification is done using discrete mosfet transistors and resistors outside the vacuum chamber.  The rise time and fall time of the voltage pulses are both 40$\pm$10~ns, with less than 5\% overshoot.  The digital pulse generator is triggered from a LabView program via a National Instruments card.  The generalised layout is shown in figure~\ref{fig:hopping_electronics_general}.  Note that this scheme to shuttle ions using switching electronics is different from the analogue voltage ramps that are typically applied to shuttle ions in RF trap arrays~\cite{ulm_pcb_trap, move_round_corner}.

\section{Results}\label{sec:hopping_results}

\begin{figure}[htbp]
\centering
\includegraphics[width=0.45\textwidth]{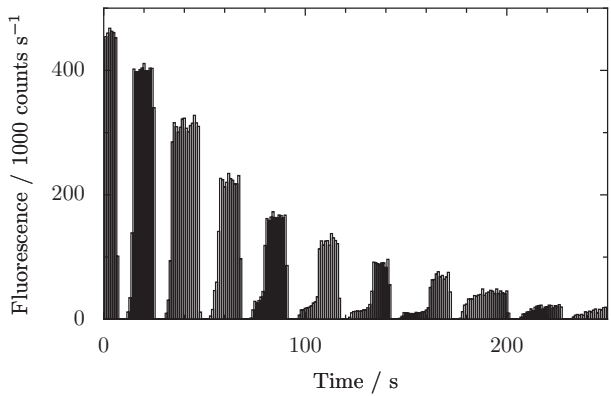}
\caption[Repeatedly shuttled ion cloud]{Fluorescence from a cloud of ions, repeatedly shuttled away from the central trap and back again.}
\label{fig:nicehops}
\end{figure}

Figure~\ref{fig:nicehops} shows the fluorescence from an initially large cloud of ions in the central trapping zone.  After a few seconds, voltage pulses are applied and the ion cloud is moved into a different trap.  The signal level drops to the background level because fluorescence is only collected from the central trap.  After a few more seconds, the reverse set of voltage pulses are applied, and roughly 75\% of the signal returns.  As the process is repeated, ions are lost on each journey, but some ions still remain after 20 shuttles (10 return trips).  The voltages used were $(V_A, V_B, V_C, V_D, V_E) = (20.0,12.5,2.5,-7.5,-15.7)$ volt, and $(V_A, V_B, V_C, V_D, V_E) = (-15.7,-7.5,2.5,12.5,20.0)$ volt.

An interesting feature can be seen each time a medium sized cloud is moved back into the central trap. A very small amount of signal returns instantly (in less time than one bin width).  The signal level then rises fairly sharply (within around 1-3 seconds), but plateaus out and remains quite low for up to 20 seconds, after which the signal level sharply rises again to its maximum value.  This behaviour does not seem to occur for very large clouds or for very small clouds.  A possible explanation for this is as follows: When an ion cloud is moved between traps, the cloud expands due to Coulomb repulsion and the ions become much hotter.  Some small fraction of the cloud however, will intersect the laser beam waist and are cool enough to produce fluorescence straight away.  The cyclotron and axial motions of the ions are cooled, and so the signal level rapidly increases.  The cloud then gradually shrinks in size as energy is transferred into the magnetron motion (the laser beam was offset to facilitate this effect)~\cite{winelanditano}.  The smaller and denser the cloud becomes, the more strongly it interacts with the laser beam, so there is a runaway effect causing the final signal increase to be rapid.

\begin{figure}[htbp]
\centering
\includegraphics[width=0.45\textwidth]{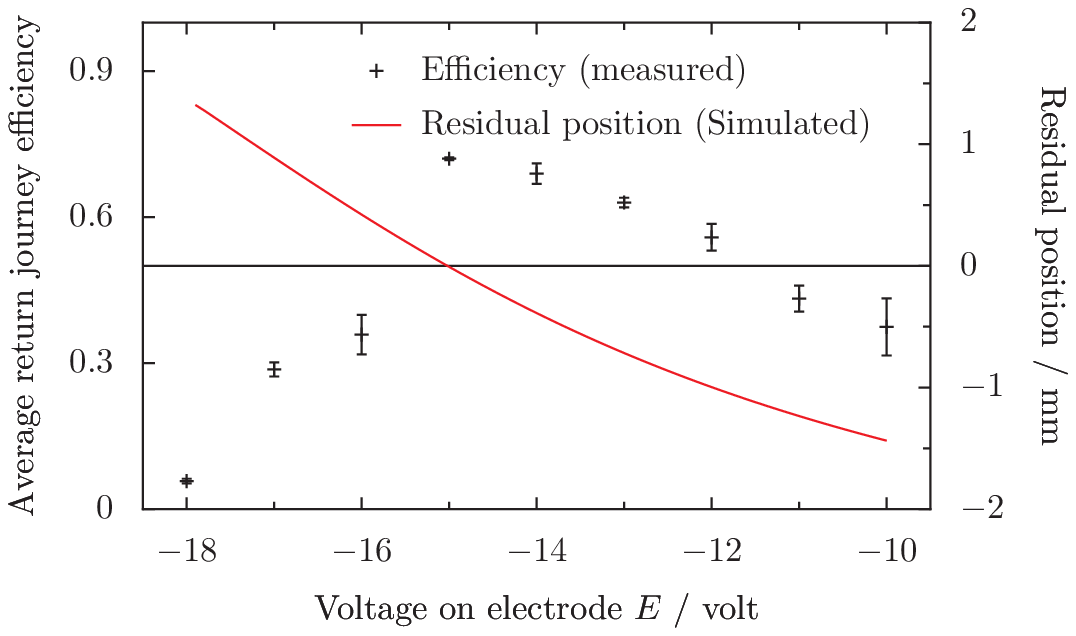}
\caption[Ion shuttling efficiency]{Efficiency of the return trip shuttling procedure, for various voltages.}
\label{fig:hopping_efficiency}
\end{figure}

The final signal level after each round trip, plotted as a function of the number of shuttles, gives a good fit to a decaying exponential.  The free parameter of this fit gives the efficiency of the shuttling procedure.  Referring back to figure~\ref{fig:displacement_vs_voltage}, changing one of the voltages away from its ideal value will cause an ion to land away from the centre of the target trap.  This was checked experimentally, altering one of the voltages and observing the transfer efficiency.  The results are shown in figure~\ref{fig:hopping_efficiency}, where the simulated residual displacement is also shown.  Two shuttling sequences were performed for each data point.  The error bar is simply the difference between the two.  It can be seen that the measured efficiency does indeed peak where the simulation predicts the optimal voltage.

Although single ions have been trapped and laser cooled for extended periods of time in this trap, single ions have not yet been reliably shuttled.  To attempt to push the shuttling efficiency up above 75\%, two improvements are being made made: A new electronics system is being built, with faster rise and fall times of $\sim$10~ns.  Secondly, coils are being wound to produce a small magnetic field perpendicular to the main field, thus allowing the angle of the field to be finely adjustable.  We will report the results of these improvements in our setup in future.

\section{Outlook and Comparison With Other Work}

This work can be compared to similar work performed using segmented RF traps.  Huber \textit{et al.} report an efficiency of 99.0\% for transporting single ions over a distance of 2~mm and back in a round trip time of 20~$\mu$s, using a segmented PCB based linear Paul trap~\cite{ulm_pcb_trap}.  Hensinger \textit{et al.} report 100\% efficiency (881 out 882 attempts) for transporting ions round a T-junction corner in 30~$\mu$s in a trap made from gold and alumina~\cite{move_round_corner}.

One potential advantage of our system is that the shuttling time is lower by about an order of magnitude (using a 1 tesla field).  If a higher field (or a lighter ion) is used, sub $\mu$s transport times would be possible~\footnote{On the other hand, future generations of smaller, tighter RF traps would also allow reliable shuttling at faster speed.  An 80\% efficiency for a single trip performed in just 3~$\mu$s was reported in~\cite{ulm_pcb_trap}.}.

Another clear advantage of our system is that the switching of electrode voltages is relatively simple.  Transport of ions between trapping zones in an RF trap has so far required a much more complicated set of precisely controlled analogue voltage ramps to be applied to the various electrodes.  

For the trap described in this paper, the best set of shuttling voltages was an ascending series over all five electrodes.  However it may be possible to build an even simpler trap with just three electrodes: a row of endcaps, and two ring type electrodes.  The shuttling could be performed by pulsing just one of the ring electrodes, and pulsing the other one to go in the other direction.  Although this goes somewhat against the simulation results shown above, the electronics would be greatly simplified.  No negative voltages or tri-state switches would be required.  If a similar but smaller trap is built, then the required shuttling voltages would also be smaller.  If the trap was designed such that the shuttling voltages were $\sim$5~V, then the pulses could be produced directly by high performance TTL, CMOS or ECL devices.  The shuttling voltage would be tuned by tuning the power supply of the fast logic device itself.  This would not only simplify the electronics, it would also allow the remarkable performance of modern digital ICs to be directly utilised.

A further advantage of the scalable Penning trap concept is seen when considering the transportation of ions round corners.  Although a near perfect efficiency is reported by Hensinger \textit{et al.}, they state that in their simulations, an ion acquires about 1.0~eV of kinetic energy during a corner-turning protocol~\cite{move_round_corner}.  Real ions in our system certainly gain some energy after shuttling, but at least in principle the gain in energy can be extremely small.  Moving an ion round a corner in a multiple Penning trap would involve nothing more than performing two regular cycloid loops at right angles to each other.  Alternatively, moving around a corner could mean shuttling along the trap axis between two axially aligned traps (using standard techniques), followed by a cycloid loop perpendicular to the magnetic field.  

The electrodes of a future trap should be designed in such a way as to facilitate ion transport round a corner, and possibly even along all three dimensions.  Building a three dimensional (or even just two dimensional) trap array, with the ability to move ions in a controlled manner between each trapping region would certainly be a challenging technical feat, but the results presented here have demonstrated the first proof of principle of the scalable Penning trap.  

\begin{acknowledgments}

This work is supported through the EU integrated project SCALA, and by the Engineering and Physical Sciences Research Council (EPSRC) of the UK.~~SA acknowledges a fellowship from the Commonwealth Scholarships Commission, U.K.

\end{acknowledgments}

\bibliography{bib}

\end{document}